\documentclass[vecphys]{svmult}		

\usepackage{makeidx}         
\usepackage{graphicx}        
\usepackage{epsfig}          
\usepackage{multicol}        
\usepackage[bottom]{footmisc}

\usepackage{color}
\usepackage{amsmath}
\usepackage{amssymb}
\usepackage[normalem]{ulem}

\makeindex             
\begin{document}

\def\jcmindex#1{\index{#1}}
\def\myidxeffect#1{{\bf\large #1}}

\title*{The Continuing Story of the Wobbling Kink 
}
\titlerunning{Wobbling Kink of the $\phi^4$}
\author{
Igor Barashenkov\inst{1,2} 
}

\institute{
University of Cape Town and
National Institute for Theoretical Physics, Western Cape, South Africa \\ \texttt{Igor.Barashenkov@uct.ac.za}
\and  Joint Institute for Nuclear Research, Dubna, Russia}

\maketitle
\abstract
{The wobbling kink is the soliton of  the $\phi^4$ model with
 an excited internal mode.
We outline an asymptotic construction of this particle-like solution that takes  into account the coexistence of 
several space and time scales. 
The breakdown of the asymptotic
expansion at large distances is prevented by 
introducing 
 the long-range variables
 ``untied"  from the short-range oscillations.
 We formulate a quantitative theory for the fading of the kink's wobbling due to the second-harmonic radiation,
 explain the wobbling mode's longevity 
and discuss ways to compensate  the radiation losses. 
The compensation is achieved by the  spatially uniform driving of the kink, external or parametric, at a variety of 
 resonant frequencies.
For the given value of the driving strength, the largest amplitude of the kink's oscillations is sustained by the 
{\it parametric\/} pumping --- at its natural wobbling frequency. This type of forcing also
produces the widest Arnold tongue in the ``driving strength versus driving frequency" parameter 
 plane. As for the  {\it external\/} driver with the same frequency, it  brings about 
an interesting rack and pinion mechanism
that converts the energy of external oscillation to the translational motion of the kink. 
}

\section{Prologue}

An unlikely insomniac wandering into the Dubna computer centre on one of those freezing nights in the winter of 1975,
would invariably see the same slim figure 
rushing among
mainframe  dashboards,
magnetic tape recorders
and card perforation devices.
A houndstooth blazer favoured by the jazzmen of the time,   a  ginger beard and a trademark cigarette holder ---  with
some amazement, the passer-by
would  recognise  Boris (``Bob") Getmanov, a pianist and 
a popular character  of the local music scene. This time, experimenting with  harmonies of nonlinear waves.

The object that inspired Bob's syncopations, was  a $\phi^4$ kink with a nonlinearly excited internal mode --- 
something he interpreted as a bound state of three kinks and called a musical term \jcmindex{\myidxeffect{T}!tritone}
 {\it tritone\/} \cite{BG4}.
In this note we review the continuation of the {\it tritone\/} story --- the line of research started with  blue sky
experiments of an artist captivated by  mysteries of the nonlinear world.

  \section{Wobbling mode}

 Getmanov's numerical experiments were motivated by similarities between 
   the $\phi^4$-theory
and the sine-Gordon model
 --- arguably, the  two simplest Lorentz-invariant nonlinear PDEs. 
While the $\phi^4$ and sine-Gordon have so much in common, there is an important difference
between the two systems. The sine-Gordon is completely integrable
whereas the $\phi^4$ model is not. Many a theorist tried to detect at least some remnants of
integrability among the properties of the $\phi^4$ kinks
--- and Getmanov with his computer codes was part of that gold rush --- but only to find more and more deviations from  the exact rules set by the sine-Gordon template.

  \vspace*{-7mm}
 \begin{figure}
\center
 \includegraphics[width=0.7\linewidth]{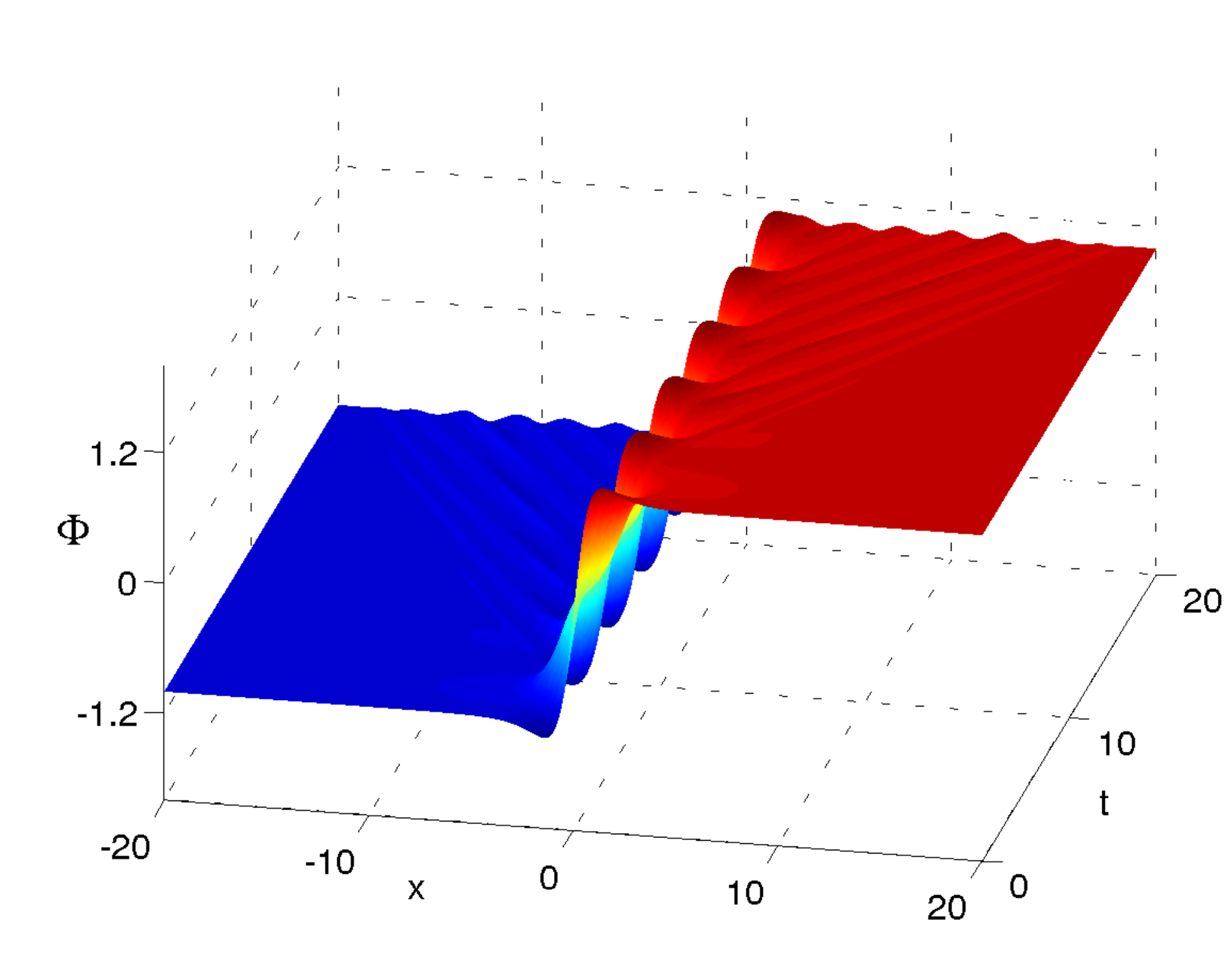}
\caption{The wobbling kink: a  kink with an activated internal mode. This solution was obtained by the 
numerical simulation of the equation (\ref{phi4}) with the initial conditions $\phi_t(x,0)=0$ and
$\phi(x,0)= \tanh x+ 2 a  \tanh x  \, \mathrm{sech} \, x$, where $a=0.3$.
}
\label{wobble}       
\end{figure} 

One attribute that  makes the $\phi^4$ kink particularly different from
its  sine-Gordon twin,  is 
the availability of an internal, or shape, mode.  \jcmindex{\myidxeffect{M}!internal mode}
(See Fig \ref{wobble}.)      
This oscillatory  degree of freedom may store energy and release it periodically --- giving rise to resonances in the kink-antikink
 \cite{KA,Belova,Uspekhi} and  kink-impurity interactions \cite{Uspekhi,FKV}, as well as
stimulating kink-antikink pair production \cite{MM,Rom2}. The internal mode 
serves as the cause of the kink's
 counter-intuitive responses to spatially-uniform time-periodic forcing  \cite{Sukstanskii,OB}  and  brings about its
 quasiperiodic velocity oscillations when the kink is set to propagate in 
 a periodic substrate potential \cite{Konotop}. 
 The excitation of the shape mode also provides a mechanism for the loss of energy and deceleration of the kink moving in a random medium \cite{Malomed}. 

\begin{figure}[t]
\center
 \includegraphics[width=0.7\linewidth]{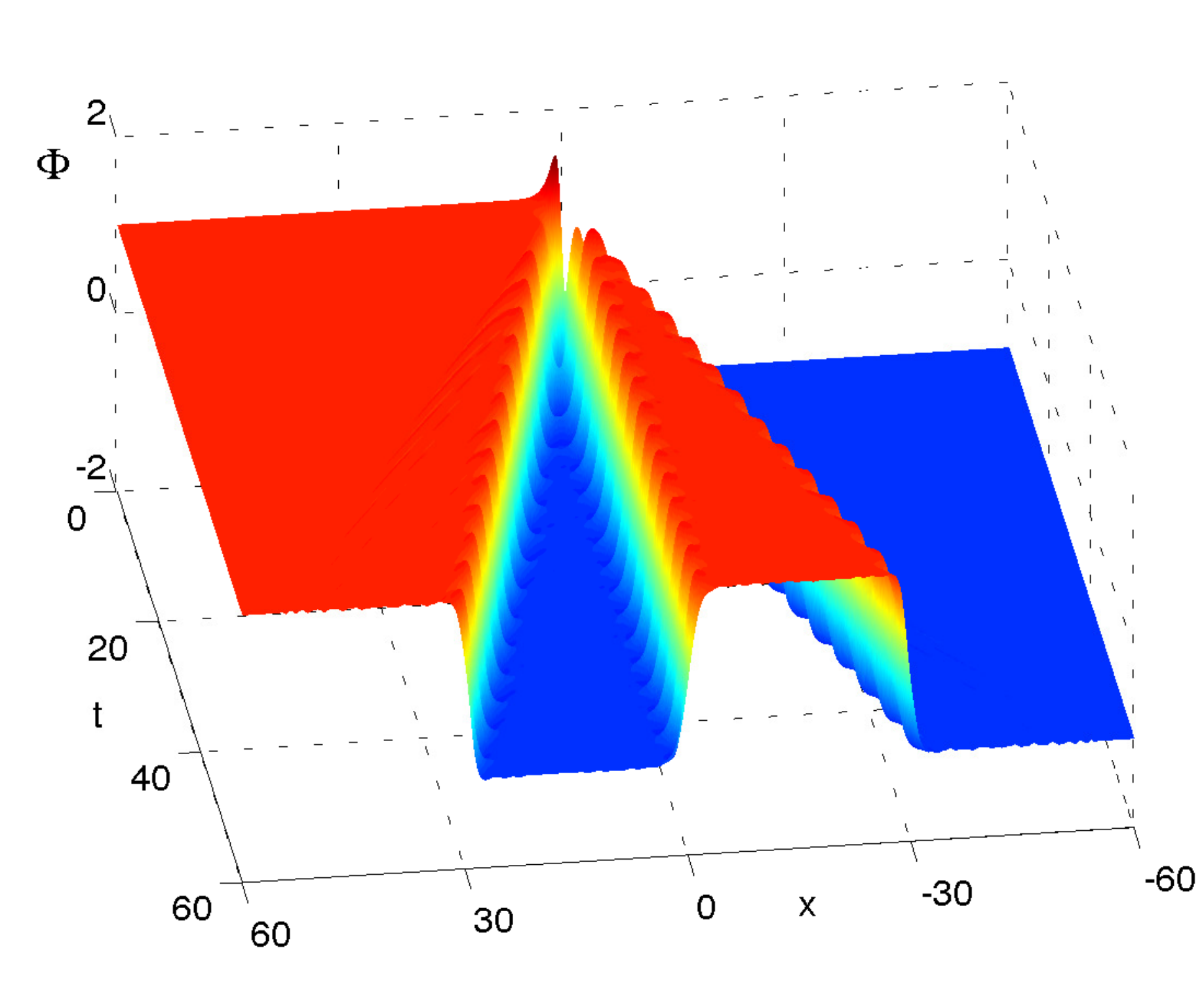}
\caption{ Evolution of the same initial condition as in Fig \ref{wobble} but with $a=0.6$.
An attempt to excite the wobbling mode with a  large amplitude  results in the emission of a kink-antikink pair. 
(Note that  time flows back to front in this figure.)
}
\label{breakup}       
\end{figure}

In his simulations, Getmanov observed an amazing longevity of  the kink's  large-amplitude  oscillations \cite{BG4}. The kink seemed  to be a comfortable location for the nonlinearity and dispersion to remain 
 balanced
in a high-energy excitation.
Getmanov regarded the nonlinear excitation of the kink as a metastable bound state of {\it three\/}  kinks --- for this object was a product of a symmetric
collision of two kinks and an antikink. 
(For more recent simulations of the bound-state formation, see \cite{Belova}.)
This interpretation is consistent with the spontaneous production  of a kink-antikink pair when the amplitude of excitation exceeds a certain threshold;
see Fig \ref{breakup}.

Getmanov's report of his {\it tritone\/} \cite{BG4} 
was eclipsed by the storm of hype around the concurrent discovery of three-dimensional  {\it pulsons\/}  --- in the same equation \cite{pulsons}.
A more sustainable wave of interest starts forming when the one-dimensional $\phi^4$ theory,
\begin{equation}
\label{phi4}
 \frac12 \phi_{tt}-    \frac12 \phi_{xx} -  \phi +  \phi^3 = 0,
\end{equation}
 was put forward
as a model for the charge-density wave materials \cite{CDW}. It has become
clear that   tritones should contribute to all characteristics of the material --- alongside kinks and breathers.
Rice and Mele described the tritone variationally, using the kink's width as the dynamical variable \cite{Rice1,Rice2}.
Segur  tried
to construct this particle-like solution as a regular perturbation expansion in powers of the oscillation amplitude \cite{Segur}. 
He determined the first two orders of  the expansion
(linear and quadratic), noted that expansion should become nonuniform at the order $\epsilon^3$
and suggested a possible way to restore the uniformity. 
To capture the antisymmetric character of oscillations,
Segur referred to the  tritone simply  as the ``wobbling kink".  \jcmindex{\myidxeffect{W}!wobbling kink}

Another interest group that has always kept an eye on the $\phi^4$ kink as the simplest topological soliton,
is the particle theorists. Unaware of  Segur's analysis,
Arod\'z and his students  developed a regular perturbation expansion 
using a polynomial approximation in the interior of the kink (see \cite{polynomial} and references therein)
and extended the  perturbation theory to  the kink embedded in $3+1$ dimensions \cite{Pelka}.
Roma\'nczukiewicz obtained an asymptotic solution for the radiation wave emitted by the wobbling mode that is initially at rest  \cite{Rom1}
and discovered kink-antiknk pair productions stimulated by the wobbling-radiation coupling \cite{Rom2}.

Segur's suggestion for the circumventing of the perturbation breakdown, was to recognise the nonlinear nature of the wobbling mode
by expanding its frequency in powers of its amplitude.
This recipe constitutes the Lindstedt method in the theory of nonlinear oscillations; 
in the wobbling kink context this approach was later implemented by Manton and Merabet \cite{MM}. 
The Manton-Merabet analysis was successful in reproducing the $t^{-1/2}$ wobbling  decay law
that had been predicted by Malomed        \cite{Malomed},  on the basis of energy considerations.  (See also \cite{Rom1}.) 

The Lindstedt method is known to be limited even when it is applied to solutions of ordinary differential equations.
It proves to be a powerful tool for the calculation of anharmonic corrections to periodic orbits --- but fails when the motion ceases to be periodic.
More importantly, the method is not well suited for the analysis of partial differential equations --- 
for it cannot handle the nonperiodic {\it spatial\/} degrees of freedom.

One alternative to the Lindstedt method is the Krylov-Bogolyubov collective-coordinate technique. This was used by Kiselev 
to obtain  solutions of the $\phi^4$ equation with a perturbed-kink initial condition posed on a characteristic line  \cite{Kiselev}. 
Although the introduction of collective coordinates allows one to recognise the hierarchy of
coexisting space and time scales in the system,
the resulting solutions lack the  explicitness and transparency of the  perturbation expansions in \cite{MM,Segur,Pelka,Rom1}. 
The physical interpretation of their  constituents is not straightforward either.

To preserve the lucidity  of  the regular expansion and, at the same time, take into account
 the coexistence of multiple space and time scales, 
  Oxtoby and the present author have designed
a singular perturbation expansion treating  different space and time variables as independent
\cite{OB,BO}.   This asymptotic construction  is reviewed in what follows. 
We also discuss
the effects of the resonant forcing of the internal mode by a variety of direct and parametric driving agents.

The emphasis of the present note is on the fundamentals of our method,  including the treatment of 
radiation, and the phenomenology of the wobbling kink's  responses to driving.
 The reader interested in mathematical detail is 
wellcome to consult the original publications \cite{OB,BO}.

The outline of this chapter  is as follows. In the next section  we explain the basics of our approach as
applied to the freely wobbling kink.  Section \ref{lifetime} considers the equation for the wobbling amplitude
and draws conclusions on the lifetime of this particle-like excitation. 
In section \ref{radiation} we formulate the asymptotic formalism for the consistent treatment of the long-range radiation. 
Section \ref{drive} is devoted to the effect of spatially-uniform temporally-resonant driving force.
Some concluding remarks are made in
 section \ref{summary}.

\section{Multiple scales:  slow times  and  long distances} 
\label{mults}

In this and the next two sections we follow  Ref \cite{BO}. 
Instead of studying  the kink travelling with the velocity $v$, 
we consider a motionless kink centred at the origin
of the reference frame that moves with the velocity  $v$ itself.
This is accomplished by the 
change of variables $(x,t) \to (\xi,\tau)$, where
$ \xi = x - \int_0^t v(t') dt'$, $\tau=t$.
The above transformation 
takes  equation (\ref{phi4})   to
\begin{equation}
\label{phi4converted}
\frac{1}{2}\phi_{\tau \tau} - v\phi_{\xi \tau} - 
\frac{v_\tau}{2} \phi_\xi - \frac{1-v^2}{2}\phi_{\xi\xi} - \phi +
\phi^3 = 0.
\end{equation}
An explicit  occurrence of the soliton velocity  in a relativistically-invariant equation
is justifiable when there are factors that can induce time-dependence of $v$
(e.g.  radiation losses)
or when the Lorentz invariance is broken by damping and driving terms.
(These are discussed in section \ref{drive}.)

We expand  $\phi$ about the kink
$\phi_0 \equiv \tanh \xi$:         \jcmindex{\myidxeffect{A}!asymptotic expansion}\begin{equation}
\label{phiexpans}
\phi = \phi_0 + \epsilon\phi_1 + \epsilon^2\phi_2 + \ldots.
\end{equation}
Here $\epsilon$ is not pegged to any small parameter of the system
(e.g. distance to  some critical value) and has the meaning of the amplitude of the kink's perturbation.
It can be chosen arbitrarily. 

We also   introduce a sequence of ``long" space
and  ``slow" time coordinates:     \jcmindex{\myidxeffect{S}!mutiple scales} 
\[
X_n \equiv \epsilon^n \xi, \quad
T_n \equiv \epsilon^n \tau, \quad n=0,1,2,....
\]
In the limit $\epsilon \to 0$,
the  $X_n$ and $T_n$ are not coupled and can be treated as independent 
variables. Consequently,  the $\xi$- and $\tau$-derivatives 
are expressible using the chain rule:
\begin{equation}
\frac{\partial}{\partial \xi} = 
\partial_0 + \epsilon \partial_1 
+ \epsilon^2 \partial_2 + \ldots, \quad 
\frac{\partial}{\partial \tau} = D_0 + \epsilon D_1 
+ \epsilon^2 D_2 + \ldots, 
  \label{chain}  \end{equation}
where 
$\partial_n   \equiv   \partial / \partial X_n$ and
$D_n   \equiv   \partial / \partial T_n$.

 We limit our analysis to the situation where the velocity is small.
Hence we write
$v = \epsilon V$  where
$V$ is of order 1. Furthermore, 
when the  wobbling amplitude $\epsilon$ is small, it is natural to expect the velocity of the kink
to vary slowly. Accordingly, $V$ is taken to be a function of slow times only:  $V=V(T_1, T_2, ...)$.

Substituting the above expansions into the $\phi^4$ equation 
(\ref{phi4converted}),  we equate coefficients of like powers of $\epsilon$.
At  the order $\epsilon^1$, we obtain the linearised equation 
\[
\frac{1}{2}D_0^2\phi_1 + {\mathcal L}\phi_1 = 0,
\]
where 
\[
{\mathcal L} = -\frac{1}{2}\partial_0^2-1+3\phi_0^2.
\]
We choose a particular solution of this equation:
\begin{equation}
\phi_1 =
 A(X_1,  X_2, \ldots; T_1, T_2, \ldots) \,
 \mathrm{sech} \, X_0\tanh X_0e^{i \omega_0 T_0} + c.c.,
 \label{phi_1}
\end{equation}
where $\omega_0=\sqrt{3}$ and $c.c.$ stands for the complex conjugate of the preceding term.
This is a wobbling mode with an undetermined 
amplitude  $A$.
The amplitude is a 
constant with respect to  $X_0$ and $T_0$ but 
will generally depend on the stretched variables $X_n$ and $T_n$   ($n=1,2, ...$).

At the second order we obtain 
\begin{equation}
\frac{1}{2}D_0^2\phi_2 + {\mathcal L}\phi_2 
= F_2,
\label{qua}
\end{equation}
where $F_2$ includes terms proportional to $e^{\pm i \omega_0 T_0}$, $e^{\pm 2 i \omega_0 T_0}$ and $e^0$:
\[
F_2= (\partial_0\partial_1 - D_0 D_1)\phi_1 
- 3\phi_0\phi_1^2+ VD_0\partial_0\phi_1 
+ \frac{1}{2}D_1V\partial_0\phi_0-\frac{1}{2}V^2\partial_0^2\phi_0.
\]
Accordingly, 
the solution  of (\ref{qua})
consists of a sum of three harmonics:
\begin{eqnarray}
\phi_2 = \varphi_2^{(0)} + \varphi_2^{(1)}e^{i\omega_0 T_0} + c.c. 
+ \varphi_2^{(2)}e^{2i\omega_0 T_0} + c.c.,
\label{phi2m}
\end{eqnarray}
where the coefficients $\varphi_2^{(0)}$, $\varphi_2^{(1)}$ and
$\varphi_2^{(2)}$ are determined to be
\begin{equation}
\varphi_2^{(0)} = 2 |A|^2 \mathrm{sech}^2X_0 \tanh X_0 
+ \left( \frac{V^2}{2}-3|A|^2 \right)
X_0   \,   \mathrm{sech}^2X_0, 
\label{varphi2}
\end{equation}
\begin{equation}
\label{phi2firstharm}
\varphi_2^{(1)} = -(\partial_1A + i\omega_0 VA)X_0\mathrm{sech} X_0\tanh X_0,
\end{equation}
and
\begin{equation} 
\label{wrongsoln}
\varphi_2^{(2)} = A^2f_1(X_0),
\end{equation} 
with
\begin{eqnarray}
f_1(X_0) =     \frac{1}{8} \left\{ 6\tanh X_0 \mathrm{sech}^2 X_0 
+ (2+ik_0 \tanh X_0+ \mathrm{sech}^2 X_0)     \right.                  \nonumber  \\            \times 
[ J^*(X_0)-J_\infty] e^{ik_0 X_0}
+ (2 -ik_0 \tanh X_0+ \mathrm{sech}^2 X_0)J(X_0) e^{-ik_0X_0} \big\}.
\label{f1}
\end{eqnarray}
 In the  expression for $f_1$,  $k_0=\sqrt 8$ and we have introduced the notations
\begin{equation}
J(X_0) = \int_{-\infty}^{X_0} e^{ik_0\xi} \mathrm{sech}^2 \xi \; d\xi,
\quad 
J_\infty=J(\infty).
\label{JJ}
\end{equation}

The odd function  $f_1(X_0)$
gives the short-range structure of the second-harmonic radiation.  
The particular solution (\ref{wrongsoln})-(\ref{f1})  corresponds to  the {\it outgoing\/}  radiation:
\begin{eqnarray}
             \varphi_2^{(2)}  e^{2i\omega_0 T_0}   &    \to    &  \frac{       2-ik_0        }{8}     J_\infty   \, A^2 
e^{i(2\omega_0 T_0- k_0X_0)}  \    
 \mbox{as}  \ X_0 \to \infty; \nonumber \\  
  \varphi_2^{(2)}  e^{2i\omega_0 T_0}   &
  \to   &   - \frac{2-ik_0}{8}   J_\infty    \,  A^2 e^{i(2 \omega_0 T_0 +k_0X_0)}  \    
 \mbox{as}  \ X_0 \to -\infty.
 \label{asrad}
 \end{eqnarray}

To obtain the coefficients (\ref{varphi2})  and (\ref{phi2firstharm})
 we had to impose  the constraints  $D_1V = 0$ and $D_1A = 0$.
These were necessary to make sure that $\varphi_2^{(0)}$ and $\varphi_2^{(2)}$ 
remain bounded as $|X_0| \to \infty$.

The $\epsilon^2$-correction to the first-harmonic coefficient, function    (\ref{phi2firstharm}), 
decays to zero as $|X_0| \to \infty$; hence the term $\varphi_2^{(1)} e^{i \omega_0 T_0}$ in 
(\ref{phi2m}) is not secular. However,
the  function    (\ref{phi2firstharm})
becomes greater than the  $\epsilon^1$-coefficient (\ref{phi_1}) once $X_0$ has grown large enough.
This contradicts our implicit assumption that all coefficients $\phi_n$  remain of the same order:  $\phi_{n+1}/ \phi_n=O(1)$
for all $X_0$.
Accordingly,  the term  $\varphi_2^{(1)} e^{i \omega_0 T_0}$ with $\varphi_2^{(1)}$  in   (\ref{phi2firstharm})
is referred to as {\it quasi\/}-secular. 
To eliminate the resulting nonuniformity in the asymptotic expansion, we have to set
\begin{equation}
\partial_1 A + i\omega_0 VA = 0.
\label{p1A}
\end{equation}
This equation will play a role in what follows.

Another quasi-secular term is the term proportional to  $X_0 \, \mathrm{sech}^2 \, X_0$ in the equation 
(\ref{varphi2}). This function  is nothing but  the derivative of $\tanh (\kappa X_0)$  with respect to  $\kappa$.
Hence we can eliminate the quasi-secular term simply by
  incorporating  the coefficient $\frac12 V^2-3|A|^2$  
  in the variable width of the kink \cite{BO}.

\section{The wobbling mode's lifetime}
\label{lifetime}

Proceeding to the order $\epsilon^3$ we obtain an equation
\[
\frac{1}{2}D_0^2\phi_3 + {\mathcal L}\phi_3 =F_3,
\]
where
\begin{eqnarray*}
F_3 = (\partial_0\partial_1 - D_0 D_1)\phi_2 + (\partial_0\partial_2 - D_0 D_2)\phi_1 
+ \frac{1}{2}(\partial_1^2-D_1^2)\phi_1 -\phi_1^3 \\ - 6\phi_0\phi_1\phi_2 + VD_0\partial_0\phi_2 
+VD_0\partial_1\phi_1 
+ VD_1\partial_0\phi_1 
+ \frac{1}{2}D_2V\partial_0\phi_0 -\frac{1}{2}V^2\partial_0^2\phi_1.
\end{eqnarray*}

The solvability condition for
the zeroth harmonic in the above equation 
reads $D_2V = 0$.
Taken together with the previously obtained $D_1V = 0$,
this implies that the kink's 
 velocity remains constant --- at least 
up to times $\tau \sim \epsilon^{-3}$. 
This conclusion was not entirely obvious beforehand. What this result implies is that 
 the radiation does not break the Lorentz invariance of the equation.  The wobbling kink and its stationary radiation 
 form a single entity that can be Lorentz-transformed from one coordinate frame to another --- just like the ``bare" kink alone.

The solvability condition for 
the first harmonic has the form of  a nonlinear ordinary differential equation:
\begin{equation}
\label{freeamp}
 i \frac{2  \omega_0 }{3} D_2A + \zeta |A|^2A - V^2A = 0,
\end{equation}
where the complex coefficient $\zeta$ is given by 
\begin{equation}
\zeta =12
\int_{0}^{\infty} \mathrm{sech}^2X_0 \tanh^3  X_0   \Big[\frac{5}{2}     \mathrm{sech} ^2X_0\tanh
X_0 
 -3X_0\mathrm{sech}^2X_0 + f_1(X_0)\Big]\,dX_0,
   \label{ze} \end{equation}
   with $f_1$ as in (\ref{f1}). 
The imaginary
part of $\zeta$ admits an explicit expression,
\[
\zeta_I = \frac{3\pi^2k_0}{\sinh^2 \left(\pi k_0/2 \right)}
=  0.046,
\]
while the real part can be evaluated numerically:
$\zeta_R = -0.85$.

Adding the equation (\ref{freeamp}) multiplied by $\epsilon^3$ and  the equation
(\ref{p1A}) multiplied by $-\frac23 i \omega_0 \epsilon v$, 
we obtain \cite{BO}
\begin{equation}
\label{mainampeqfree}
i \frac{ 2\omega_0}{3} a_t + \zeta  |a|^2a + 
 v^2a +   O(|a|^5)=0,
\end{equation}
where we have introduced  the ``unscaled"  amplitude of 
the wobbling mode
$a= \epsilon  A $
and reverted to the original kink's velocity $v = \epsilon V$.
We have also  used that  $A_t= \epsilon (D_1-v  \partial_1)A + \epsilon^2 D_2 A +
O( {\epsilon^3})$.
The advantage of the equation (\ref{mainampeqfree}) over (\ref{freeamp}) 
is  that the expression (\ref{mainampeqfree})
remains valid for all times from $t=0$ to $t \sim \epsilon^{-2}$
whereas the equation (\ref{freeamp})  governs the evolution only on long time intervals.

According to (\ref{mainampeqfree}), the amplitude
 will undergo a monotonic
decay:
\begin{equation}
 \label{decay_rate}
|a(t)|^2 = \frac{|a(0)|^2} 
{1+ \omega_0 \zeta_I \, |a(0)|^2t }.
\end{equation}
This equation was originally derived by the Lindstedt method in \cite{MM} and using energy considerations in \cite{MM,Rom1}.

The  smallness of $\zeta_I$  can be deduced from
equation (\ref{ze}) even without performing the exact integration. Indeed,
the imaginary part of the integral (\ref{ze})  is of the form 
\begin{equation}
\zeta_I = \int_0^\infty (  F_0 \cos k \xi + G_0 \sin k \xi)  \,d\xi,
\label{AcBs}
\end{equation}
where $k>1$  and
the real functions $ F_0(\xi$) and  $G_0(\xi)$ are  even and odd, respectively.
The functions $F_0(\xi)$, $G_0(\xi)$ and all their derivatives are bounded on $(0,\infty)$ and
 decay to zero as $\xi \to \infty$.
A repeated integration by parts  gives
\begin{equation}
\zeta_I(k)= \sum_{n=0}^N    \frac{G_n(0)}{k^{n+1}}+ O\left( \frac{1}{k^{N+1}} \right),
\label{series}
\end{equation}
where
\[
G_{n+1}(\xi) = - \frac{d F_n}{d\xi}, \quad F_{n+1}(\xi)= \frac{dG_n}{d \xi} \quad   (n=0,1,2,...).
\]
Because of the evenness of $F_0(\xi)$ and oddness of $G_0(\xi)$, all coefficients in the  series (\ref{series}) are zero.
Therefore the integral (\ref{AcBs})   is
smaller than any positive power of $1/k$ --- that is, it is {\it exponentially\/} small as $k \to \infty$.
As a result, even with a moderate value of $k$, $k=k_0=\sqrt 8$, 
 we have $\zeta_I$ 
below $0.05$.

The fact that the  decay rate $\zeta_I$  is an exponentially decreasing function of the 
radiation wavenumber, has a simple physical explanation.  The  decay rate of the wobbling mode's energy is determined by  the energy flux which, in turn, is proportional 
to the square of the radiation wave amplitude. On the other hand,
the amplitude of radiation from any localised oscillatory mode (a
localised external source, an impurity or internal mode) is an exponentially decreasing
function of $k$.
In particular, the 
 amplitude of the second-harmonic radiation excited by the wobbling mode 
includes the factor $J_\infty$ --- see equation (\ref{asrad}). This factor is evaluated to
\[
J_\infty=    \int_{-\infty}^{\infty} e^{ik_0\xi} \mathrm{sech}^2 \xi \; d\xi =            \frac{\pi k_0}{\sinh(\pi k_0/2)}.
\]
Accordingly, if $k_0$ were allowed to grow rather than being set to $\sqrt 8$, the energy flux would drop  in proportion to 
 $e^{-\pi k_0}$.

Thus  the longevity of the wobbling mode is due to 
 the wavelength of the second-harmonic radiation being several times shorter than the effective width of the kink  (more specifically,
$\pi k_0 $ being about nine times greater  than $1$).
 As a result, the decay rate $\zeta_I$ ends up having a tiny factor of $e^{-9}$.

\section{Radiation from a distant kink}
\label{radiation}

 \jcmindex{\myidxeffect{R}!radiation waves}

The above   treatment of the radiation  is limited in two ways. Firstly, although higher terms 
in the expansion (\ref{phiexpans})  were assumed to be smaller than the lower ones,
that is, $\epsilon^{n+1} \phi_{n+1} / \epsilon^n \phi_n \to 0$ as $\epsilon \to 0$,
 the coefficient $\phi_1$ becomes exponentially small 
 while $\phi_2$ remains of order 1 as $|X_0|$ grows.
This means that our construction is only consistent for not too large values of $|X_0|$. 

Second, the asymptotic expansion (\ref{phiexpans}) with coefficients (\ref{phi_1}) and (\ref{phi2m})
describes a steadily oscillating kink but cannot account for any perturbations propagating in the system.
Let, for simplicity, $V=0$ and choose some initial condition for the amplitude $A$:
$A=A_0$ at $T_2=0$.
 Then $\phi$ is equal to
\[
\phi= -1 - \epsilon^2  \frac{2-ik}{8} J_\infty      \,   A_0^2      \,        e^{ik_0 X_0}
\]
for {\it all\/} sufficiently large negative $X_0$ (and to the negative of this expression for all large positive $X_0$). 
This creates an impression  that 
the initial perturbation  has travelled a large distance instantaneously --- while in actual fact 
the asymptotic solution that we have constructed describes a (slowly relaxing) stationary structure and 
cannot capture transients or perturbations.

To design a formalism for the propagation of  nonstationary waves  we perform a Lorentz transformation  to the reference frame where $v=0$.
(As we have explained in the previous section, the radiation from the kink does not break the Lorentz invariance.)
Consider large positive $X_0$ 
and expand the field as in
 \begin{equation}
 \phi=1+ \epsilon^2 \phi_2+ \epsilon^4
\phi_4+\dots.
\label{oexp}
\end{equation} 
In a similar way,  we let 
\begin{equation}
\phi=-1+ \epsilon^2 \phi_2+ \epsilon^4
\phi_4+\dots
\label{oexp2}
\end{equation}
for large negative $X_0$.
Substituting these, together with the derivative expansions 
(\ref{chain}),
 in equation (\ref{phi4}), 
 the order $\epsilon^2$ gives
 \begin{equation}
 \phi_2= {\cal J} B_+ e^{i(\omega_+ T_0-k_+ X_0)}+c.c.,
 \quad X_0>0,
 \label{Bp}
 \end{equation} 
 and 
 \[
 \phi_2= -{\cal J} B_- e^{i(\omega_- T_0-k_- X_0)}+c.c.,
 \quad
 X_0<0.
\]
Here 
$\omega_{\pm}=\sqrt{ k_{\pm}^2+4}$,
and the amplitudes $B_\pm$ are functions
of the stretched  coordinates: $B_\pm=B_\pm(X_1,...; T_1,...)$. 
The coefficient ${\cal J}$ will be chosen at a later stage
and  the negative sign in front of $B_-$   was also introduced for later convenience.

Consider a point $X_0= \frac12 \ln \epsilon^{-1}$. Sending  $\epsilon \to 0$
we have $X_0 \to \infty$, so that the ``outer" expansion (\ref{oexp}) with $\phi_2$ as in  (\ref{Bp}) is valid.
 On the other hand, 
the ratio $(\epsilon^2 \phi_2)/(\epsilon \phi_1)$ with $\phi_1$ and $\phi_2$ as in 
(\ref{phi_1}) and (\ref{phi2m}),  is $O(\epsilon^{1/2})$; 
hence the ``inner" expansion 
 (\ref{phiexpans}) remains uniform at the chosen point.

The corresponding stretched coordinates $X_1=(\epsilon/2) \ln \epsilon^{-1}$, 
$X_2 =(\epsilon^2/2) \ln \epsilon^{-1}$, ..., 
satisfy $X_1, X_2, ... \to 0$ as $\epsilon \to 0$.  Consequently,   the coefficient  $B_+$ in 
 (\ref{Bp}) has zero spatial arguments:
$B_+=B_+(0,0,...;T_1,T_2, ...)$.
In a similar way, the amplitude $A^2$ in  (\ref{phi2m}) is  $A^2(0,0, ...; T_2, T_3, ...)$.
Choosing ${\cal J}= \frac18 (2-i k_0) J_\infty$ and
equating (\ref{Bp}) to (\ref{phi2m}), we obtain 
$\omega_+=2\omega_0$, $k_+= k_0$, and
\begin{equation}
B_+(0,0,...;T_1,T_2,...)= A^2(0,0,...;T_2,T_3,...).
\label{match} 
\end{equation}
A similar matching at the point $X_0= -\frac12 \ln \epsilon^{-1}$ leads to
\begin{equation}
B_-(0,0,...;T_1,T_2,...)= A^2(0,0,...;T_2,T_3,...).
\label{match2} 
\end{equation}

The solvability condition for the nonhomogeneous equation arising at
 the order $\epsilon^3$, gives
a pair of  linear transport equations
 \begin{eqnarray}
D_1{ B}_+ +  c_0 \partial_1{ B}_+ =0, \quad X_1 >0, 
\label{transport1}\\
D_1{ B}_- -  c_0 \partial_1{ B}_- =0, \quad X_1 <0,
\label{transport2}
\end{eqnarray}
where  $c_0=k_0/(2 \omega_0)$.
Equation (\ref{transport1}) should be solved under the boundary condition (\ref{match})
while solutions of (\ref{transport2}) should satisfy the condition (\ref{match2}).

Solutions of the equation (\ref{transport1}) propagate along the characteristic lines
$X_1=c_0 T_1 +\eta$
where $\eta$   is a parameter  ($-\infty<  \eta < \infty$). In a similar way, solutions of (\ref{transport2})
travel along the characteristics $X_1=-c_0 T_1 +{ \tilde \eta}$.
The velocity $c_0$ is the group velocity for wavepackets of second-harmonic radiation 
centred on the wavenumber $k_0$; as one can readily check,
$c_0=\left. (d \omega/dk) \right|_{k=k_0}$, where
$\omega=\sqrt{4+k^2}$.
This velocity is of course smaller than the speed of light: $c_0<1$. 

If the amplitude $B_+$ satisfies the initial condition
\[
 B_+(X_1,0)= \beta(X_1), \quad  X_1>0  \]
at the moment $T_1=0$, its subsequent evolution in the  region $X_1>c_0  T_1$ for $T_1>0$
 is a mere translation:
\[
 B_+(X_1, T_1)=   \beta(X_1-c_0T_1).
\]
In a similar way, in the  region $X_1< -c_0 T_1$  with $T_1>0$  the amplitude
$B_-$ satisfies 
\[
 B_-(X_1, T_1)=   \beta(X_1  +c_0 T_1),
\]
where $\beta(X_1)$ is the initial condition for this solution:
\[   
B_-(X_1,0)= \beta(X_1), \quad  X_1<0.  
 \]
In the sector between the rays $X_1= -c_0 T_1$  and $X_1=c_0 T_1$, the solutions are defined by the boundary conditions instead of initial ones:
\begin{eqnarray*} 
 B_+(X_1,T_1)=  \left. A^2 \right|_{X_1=T_1=0}, \quad  &  0< X_1 < c_0T_1; \\
B_-(X_1,T_1)=   \left. A^2\right|_{X_1=T_1=0}, \quad    &  -c_0T_1< X_1 <0.
\end{eqnarray*}

If we choose the initial condition $\beta(X_1)=0$ for all $X_1$, 
the field $\phi$ will have the form  of a pair of fronts, or shock waves, propagating away from the origin. 
For $|X_1|> c_0T_1$, the amplitudes  will be zero, $B_\pm=0$,  whereas for $|X_1|< c_0T_1$, these will assume nonzero constant values:
$B_\pm (X_1,T_1)=  \left. A^2 \right|_{X_1=T_1=0}$. 
In agreement with one's physical intuition, the wobbling kink only influences the adjacent domain 
$|X_1| <c_0 T_1$.

To describe the evolution of the radiation on a longer, $T_2$, scale, one needs to derive 
one more pair of amplitude equations for $B_+$ and $B_-$. 
The  coefficient $\phi_4$ in  the expansions (\ref{oexp}) and (\ref{oexp2}) can be determined 
if the following solvability conditions are satisfied  --- in the right and left half of the $x$-line, respectively:
 \[
i(2 \omega_0 D_2+ k_0 \partial_2) B_\pm 
+\frac12 (D_1^2-\partial_1^2) B_\pm 
=0.
\]
Eliminating $D_1 B_\pm$ using  (\ref{transport1}) and (\ref{transport2}),  the above equations are transformed to
\begin{equation}
iD_2 B_\pm \pm i c_0 \partial_2 B_\pm
-
\frac{\omega_{kk}}{2} \partial_1^2 B_\pm=0.
\label{2nd_cor} 
\end{equation}
Here
\[
\omega_{kk} \equiv \left. \frac{d^2 \omega}{dk^2} \right|_{k_0}=
\frac{4 \omega_0^2-k_0^2}{8 \omega_0^3}
\]
 is the dispersion of
the group velocity of the radiation waves. 
Adding
(\ref{2nd_cor})  and (\ref{transport1})-(\ref{transport2}) multiplied by the appropriate powers of $\epsilon$, 
 we obtain a pair of Schr\"odinger
equations in the laboratory coordinates:
\begin{equation}
i \partial_t  B_\pm  \pm ic_0  \partial_x  B_\pm
 -\frac{\omega_{kk}}{2} \partial_x^2 B_\pm=0. 
\label{rfin} 
\end{equation}

Consider the top equation in (\ref{rfin})  on the interval ${\tilde x} <x< \infty$, with the initial condition $B_+(x,0)=0$
and the boundary conditions 
\[
B_+({\tilde x},t)= A^2(\epsilon^2 t),  \quad B_+(\infty,t)=0. 
\]
Here $\tilde x= \frac12 \ln \epsilon^{-1}$. 
The solution evolving out of this combination of 
initial and boundary conditions will describe a step-like front  propagating to the right with the
velocity $c_0$ and dispersing on the slow time scale $t \sim \epsilon^{-2}$. In a similar way, 
the bottom equation in (\ref{rfin}),  considered on the interval $-\infty<x <-\tilde x$ 
with the initial condition $B_-(x,0)=0$
and boundary conditions $B_-(-{\tilde x},t)= A^2(\epsilon^2 t)$, $B_-(-\infty,t)=0$,
will describe a slowly spreading shock wave moving to the left.

This completes the multiscale description of the freely  radiating kink \cite{BO}.
The uniformity  of the asymptotic
expansion at all scales  is secured by 
introducing 
 the long-range radiation variables
 which are related to the short-range radiations through boundary conditions but do not automatically coincide with those.

\section{Damped driven wobbling kink}
\label{drive}

The wobbling kink may serve as a stable source of radiation with a certain  fixed frequency and wavenumber.
However in order to sustain the wobbling,  energy has to be fed into the system from outside.

 A particularly efficient and uncomplicated way of pumping  energy into a kink is
 by means of a resonant driving force.
This type of  driving does not have to focus
on any particular location; the driving wave 
 may fill, indiscriminately, the entire  $x$-line. Only the object in possession of the resonant internal mode will respond to it.
Furthermore, a small driving amplitude should be sufficient  to sustain the oscillations as the free-wobbling decay rate is very low.

The two standard ways of  driving an oscillator  are  by applying an external force in synchrony with its own natural oscillations
or by varying, periodically, one of its parameters. The former is commonly called the 
{\it direct}, or {\it external}, forcing, whereas the latter goes by the name of {\it parametric\/}
 pumping. In the case of the harmonic oscillator, 
 the strongest parametric resonance 
occurs if the driving frequency is close to twice its natural frequency. 

There is an important difference between 
 driving a structureless oscillator and pumping energy into the wobbling mode of the kink.
The mode has an odd, antisymmetric, spatial structure so that the oscillation in the positive semi-axis  is out of phase with the oscillation in the negative half line.
It is  not obvious whether the energy transfer from a spatially-uniform force to this antisymmetric mode is at all possible,
and if yes --- what type of driver (direct or parametric) and what resonant frequency 
would ensure the most efficient transfer. To find answers to these questions, we will examine driving forces of both types and with several frequencies.

In addition to the driving force,
our analysis  will take into account the dissipative losses. (The dissipation is an effect that is difficult to avoid in  a physical system.)
Accordingly,  the resulting amplitude equation  will have two types of damping terms: the linear damping
accounting for losses due to friction, absorption, incomplete internal reflection of light or similar imperfections of the physical system --- and
cubic damping due to the emission of the second-harmonic radiation.

The damped-driven $\phi^4$ equation was utilised to model
  the drift of domain walls 
in  magnetically ordered crystals placed  in oscillatory magnetic
fields  \cite{Sukstanskii};
 the Brownian motion of string-like objects on a 
periodically modulated bistable
substrate \cite{Borromeo};
ratchet dynamics of kinks in a lattice of point-like inhomogeneities
\cite{Molina} and rectification in Josephson junctions
and optical lattices \cite{MQSM}. 

The first consistent treatment of the periodically forced $\phi^4$ kink was due to Sukstanskiii and Primak \cite{Sukstanskii}
who examined the joint action  of  external and parametric excitation,
at a generic driving frequency. Using   a combination of the Lindstedt method and the method of averaging, they have discovered   the kink's drift
with the velocity proportional to the product of the external and parametric driving amplitudes. It is important to note that this effect is not related to the wobbling of the kink; 
yet the drift velocity develops a peak  when the driving frequency is near the frequency of the wobbling.

The individual effect of the external  driving force was explored by Quintero, S\'anchez and Mertens using an heuristic 
collective-coordinate Ansatz and disregarding radiation
\cite{QuinteroDirect}. As in Ref \cite{Sukstanskii}, the driving frequency was generic, that is, not pegged to any particular value.
The analysis  of Quintero {\it et al\/} suggests that  a resonant energy transfer from the driver to the kink takes place when the driving frequency is close to the frequency of the wobbling and, at a higher rate,
when the driving frequency is near a half of that value. Although the resonant transfer is not quantified in the approach of Ref \cite{QuinteroDirect} and the reported collective coordinate 
behaviour is open to interpretation, the discovery of the subharmonic resonance (supported by simulations of the full PDE) identified an important direction for more detailed scrutiny.
An interesting numerical observation of  Ref \cite{QuinteroDirect} was a chaotic motion of the kink when the driving frequency is set to a half of the natural wobbling frequency.

In the subsequent publication \cite{QuinteroParametric}, Quintero, S\'anchez and Mertens extended their analysis to the kink driven parametrically  --- still at a generic frequency.
As in their previous study, the authors observed a resonant energy transfer when the driving frequency is near the natural wobbling frequency of the kink. 
On the other hand, the parametric driving would not induce any translational motion of the kink.

Below,   we examine the resonantly driven wobbling kink using our method of multiple scales.
The exposition of this section follows Ref \cite{OB}.

\subsection{Parametric driving at the natural wobbling frequency}

 \jcmindex{\myidxeffect{P}!parametric driving}

The parametric driving of the harmonic oscillator
 is known to  cause a particularly rapid (exponential) growth of the amplitude of its oscillations.
In this section we examine the   parametric driving of the  kink 
 close to its natural wobbling frequency.

 The driven $\phi^4$ equation has the form  \cite{QuinteroParametric}
\begin{equation}
\frac{1}{2}\phi_{tt}-\frac{1}{2}\phi_{xx} 
 -\phi+\phi^3 =        -\gamma \phi_t                             +  h \cos (\Omega t) \phi. 
\label{dd}
\end{equation}
Here the driving frequency $\Omega$ is  slightly detuned from $\omega_0 = \sqrt{3}$,
the linear wobbling frequency of the undriven kink: 
\[
\Omega = \omega_0(1+\rho).
\]
It is convenient to express the small detuning $\rho$ and
the amplitude of the wobbling mode in terms of the same small parameter:
\[
\rho = \epsilon^2 R,
\]
where  $R=O(1)$.

The small parameters $h$ and $\gamma$ 
 in  the equation (\ref{dd})
measure the driving strength and
 linear damping, respectively. 
It is convenient to choose the following scaling laws for these coefficients:
\[
\gamma = \epsilon^2 \Gamma, \quad
h = \epsilon^3 H, 
\]
where $\Gamma, H=O(1)$. The above choice ensures that 
 the damping and driving terms  appear
  in the amplitude
equations at the same order as the cubic nonlinearity.

A pair of  equations produced by
the multiple-scale procedure includes an equation for 
the ``unscaled" amplitude of the wobbling mode, $a= \epsilon A$ \cite{OB}:
\[
{\dot a} = -\gamma a- i \omega_0 \left(\rho   + \frac{v^2}{2}\right)  a 
  +
 \frac{i}{2} \omega_0 \zeta|a|^2a  
 - 
 i \frac{\pi}{8} \omega_0 h + O(|a|^5).
 \]
 Here  the overdot indicates differentiation with respect to $t$ and
the complex coefficient $\zeta$ has been evaluated in section \ref{lifetime}. 
 The second amplitude equation governs the velocity of the kink:
 \begin{equation}
\label{paramomegaamplitudev}
{\dot v} = -2\gamma v + O(|a|^5).
\end{equation}

 \begin{figure}
\center
\epsfig{figure=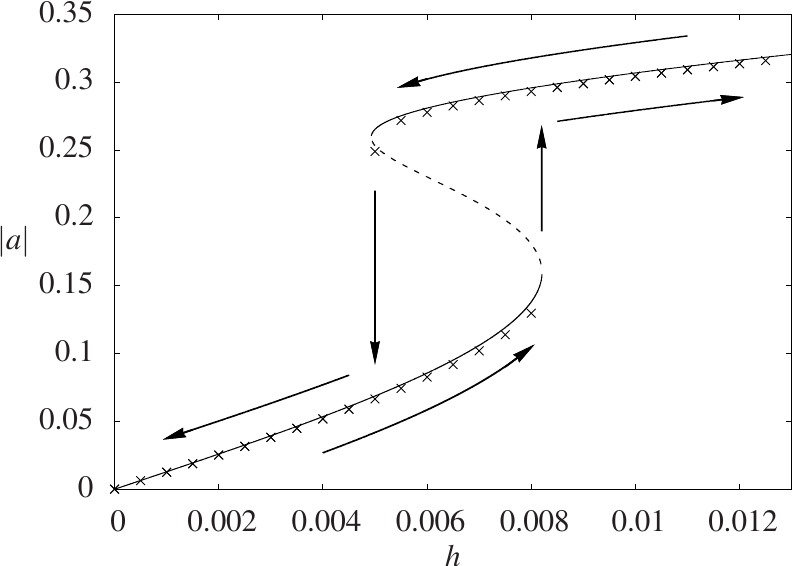,width=8cm,angle=0}
\caption{The hysteresis loop of the  $\phi^4$ kink driven, parametrically,  at the frequency close to its natural wobbling frequency.
  Continuous  curves  delineate  stable and dashed curves unstable  fixed points of the amplitude  equation (\ref{amp2}).
Crosses represent  numerical simulations of the full partial differential equation (\ref{dd}) with the same $\gamma$ and $\rho$. 
In the simulations, the driving strength $h$ was increased from
zero  in small steps and then turned  back to zero
as indicated by the arrows.
This figure was generated using $\gamma=0.01$ and $\rho=-0.03$.}
\label{fig1}       
\end{figure}

The damping term $\gamma \phi_t$ breaks the Lorentz invariance of the $\phi^4$ model (\ref{dd}). 
The kink travelling at the velocity $v$ is no longer equivalent to the kink at rest.
Instead, according to (\ref{paramomegaamplitudev}), 
the traveling kink suffers deceleration --- until $v$ becomes of order $|a|^3$. 
Thus, after an initial transient of the length $t \sim \gamma^{-1}$, the  dynamics 
is governed by a single equation for the complex wobbling amplitude:
\begin{equation}
{\dot a} + i \omega_0 \rho a - \frac{i}{2} \omega_0 \zeta|a|^2a 
= -\gamma a  - 
 i \frac{\pi}{8} \omega_0 h.
 \label{amp2}
 \end{equation}
 
 Although the underlying partial differential equation (\ref{dd})
  is driven {\it parametrically},  the forcing term in (\ref{amp2}) is characteristic for the  {\it externally\/} driven Schr\"odinger equations  (see \cite{direct} and references therein). 
The reason is that the oscillator that we are trying to pump energy to, is the kink's internal mode rather than the kink itself.
On the other hand, the leading part of the   forcing term in the right-hand side of (\ref{dd}),  $h \cos(\Omega t) \phi_0$,
involves the kink ($\phi_0$) rather than the internal mode ($\phi_1$). 
As a result,  the driver that was expected to affect a parameter of the oscillatory mode, acts as an external periodic force with regard to this mode.

 The equation (\ref{amp2}) has no attractors other than  fixed points.   Assume the damping coefficient $\gamma$  is fixed.
 When the detuning $\rho$ satisfies $\rho> \rho_0$, where
 $  \rho_0= -1.14 \gamma$,
 the dynamical system (\ref{amp2}) has a single  fixed point irrespectively of the value of $h$.
 The fixed point is stable and  attracts all trajectories.
 If the detuning is chosen in the complementary domain 
 $\rho<\rho_0$, the structure of the phase space depends on the value of $h$. 
 Namely, there are two critical values $h_+$ and $h_-$,  where $h_+<h_-$, such that 
 when $h$ is smaller than $h_+$ or larger than $h_-$,
   the system's
 phase portrait features a single fixed point --- and this point is attractive.
 The inner region $h_+<h<h_-$ is characterised by  two stable fixed points and exhibits hysteresis.
  (See Fig \ref{fig1}.)  The values  $h_+$ and $h_-$ 
  are expressible through $\rho$ and $\gamma$ \cite{OB}. 

 \jcmindex{\myidxeffect{H}!hysteresis} \jcmindex{\myidxeffect{B}!bistability}

The memory function associated with the bistability and hysteresis of the parametrically driven  wobbling kink,
endows this structure with potential applications in electronic circuits, devices based on ferromagnetism or ferroelectricity, and
charge-density wave materials.

\subsection{Parametric driving at twice the natural wobbling frequency}

The longest and widest Arnold tongue of the parametricaly
driven harmonic oscillator  corresponds to the subharmonic resonance
where the driving frequency is close to twice the natural frequency of the oscillator.
It is therefore interesting to examine this driving regime in the context of the wobbling kink.
Would the $2:1$ frequency ratio sustain a larger wobbling amplitude than the $1:1$ regime considered in the previous subsection?

The driven equation in this case has the form 
\begin{eqnarray}
\frac{1}{2}\phi_{tt}-\frac{1}{2}\phi_{xx} 
-\phi+\phi^3 =  -\gamma \phi_t
+ h\cos(2\Omega t ) \phi,
\label{pd21}
\end{eqnarray}
where, as in the previous subsection, we set
\[
\Omega = \omega_0(1+\rho).
\]
The scaling laws for the weak detuning and small driving amplitude are chosen as before:
\[
\rho=\epsilon^2 R,
 \quad \gamma= \epsilon^2 \Gamma.
\]
This time, however, it is convenient to  choose a different scaling law
 for the  small driving amplitude:
\[
h = \epsilon^2H.
\]

The resulting amplitude system is \cite{OB}
\begin{eqnarray*}
\label{param2omegaamplitudea}
{\dot a} = -\gamma a - i\omega_0  \left(         \rho +\frac{v^2}{2} \right) a + \frac{i}{2}\omega_0 \zeta |a|^2 a 
  - \frac{i}{2}  \omega_0 \sigma h a^* +
O(|a|^5), \\
{\dot v} = -2\gamma v + O(|a|^5).
\end{eqnarray*}
Here the complex coefficient $\sigma=\sigma_R+i \sigma_I$ is given by an integral
\begin{equation}
 \sigma =\int_{-\infty}^{\infty} \left[ 
\frac{1}{2} \mathrm{sech}^2 X_0 \tanh^2 X_0 
- 6   \mathrm{sech}^2 X_0 \tanh^3 X_0 f_2(X_0) \right] dX_0,
\label{sigint}
\end{equation}
where $f_2$ consists of the second-harmonic radiation and a stationary
standing wave --- both induced by the driver:
\[
f_2(X_0) = -\frac{1}{12} f_1(X_0)+ \frac{1}{24} \tanh X_0 (2 \mathrm{sech}^2 X_0-3).
\]
In the above expression, the function $f_1$  is given by the equation (\ref{f1}). The real and imaginary parts of the integral  (\ref{sigint}) are found to be
\[
\sigma_R=0.60,
\quad
\sigma_I = \frac{1}{12} \zeta_I=0.0039.
\]

\begin{figure}
\center
\includegraphics{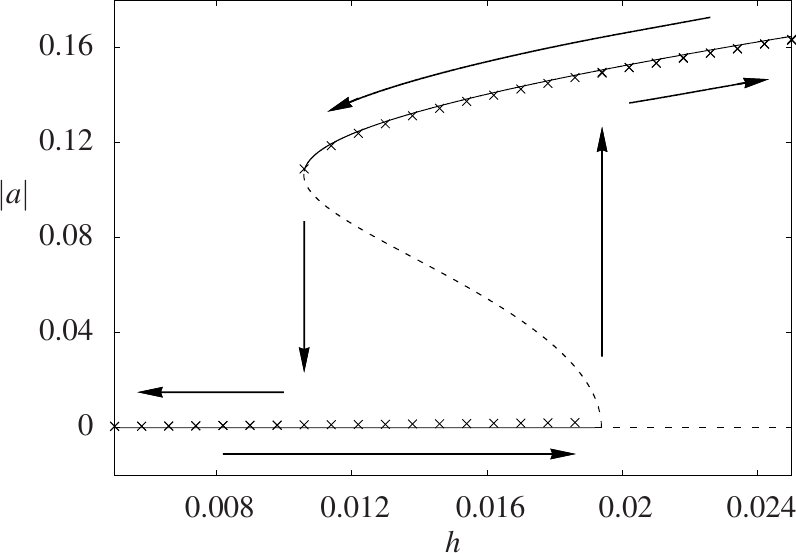}
\caption{\label{simulations21} 
The hysteresis loop of the  kink driven, parametrically, at the frequency close to twice its natural wobbling frequency.
 Continuous
and dashed 
 curves  delineate  stable and 
 unstable fixed points of the amplitude
equation (\ref{dota})  with  $\gamma=0.005$ and $\rho=-0.005$.
Crosses represent results of the  numerical simulations of the equation
(\ref{pd21}) with the same $\gamma$ and $\rho$.
The driving strength $h$ was increased from $0.005$ 
to $0.025$ in small steps and then turned back
to $0.005$ as indicated by the arrows.
 }
\end{figure}

After a transient of the length $t \sim \gamma^{-1}$, the dynamics is governed by 
a two-dimensional dynamical system
\begin{equation}
{\dot a} +  i\omega_0 \rho a- \frac{i}{2}\omega_0 \zeta |a|^2 a 
= -\gamma a 
   - \frac{1}{2}   i\omega_0 \sigma h a^*.
\label{dota}
\end{equation}
Equation  (\ref{dota}) is similar to the  equation (\ref{amp2}) from the previous section; the only difference between the two expressions 
 is the type of the driver.
Unlike equation (\ref{amp2}), the amplitude equation  
(\ref{dota}) features the parametric forcing  in its standard  Schr\"odinger form (see e.g. \cite{parametric}).

The energy-transfer mechanism in the present case is different from the mechanism operating in   the $1:1$ forcing regime. 
In the partial differential equation (\ref{pd21}),
the function $ \cos (2 \Omega t)$ in
the product $h \cos (2 \Omega t) (\phi_0+ \epsilon \phi_1+...)$, 
acts as a parametric driver for  the wobbling mode $\phi_1$.
In addition, the ``external force" $h \cos (2 \Omega t) \phi_0$
excites a standing wave with the frequency $2 \Omega$ and generates radiation at the same frequency. 
Both couple to the wobbling mode --- parametrically, via the term $\epsilon^3 \phi_0 \phi_1 \phi_2$ in the expansion of $\phi^3$. 
As a result, in the case of the equation (\ref{pd21}) we have three concurrent mechanisms at work, and all three are of parametric nature.
Combined, the three mechanisms produce the $ha^*$-term in the amplitude equation (\ref{dota}).

 Turning to the analysis of the amplitude equation, we
 assume, first, that  $\rho> \rho_0$, where
 $\rho_0   =-0.031 \gamma$. There is a critical value of the driving amplitude $h_+(\gamma,\rho)$ such that 
 the
dynamical
system (\ref{dota}) 
has two stable fixed points $a_1$ and $-a_1$  in the region $h>h_+$
and a single stable fixed point $a=0$ in the complementary region $h<h_+$.

If the frequency detuning is taken to satisfy  $\rho< \rho_0$, all trajectories flow 
to the origin for $h$ smaller than $h_-$  (where $h_-$ is another critical value expressible through $\gamma$ and $\rho$), 
 and to  one of the two nontrivial fixed points $\pm a_1$
  for large $h$ ($h>h_+$).
In the intermediate 
region $h_-< h<h_+$ the system shows a tristability between 
$a=0$ and the pair of points $a= \pm a_1$.
See Fig \ref{simulations21}.

\subsection{External subharmonic  driving}

 \jcmindex{\myidxeffect{E}!external driving}

Proceeding to the analysis of the direct driving force, we start with the case where the driving frequency 
is approximately a half of  the natural
wobbling frequency of the kink \cite{QuinteroDirect}:
\begin{equation}
\frac{1}{2}\phi_{tt}-\frac{1}{2}\phi_{xx}  - \phi + \phi^3 = - \gamma \phi_t  +
h\cos \left(\frac{\Omega}{2} t \right).
\label{dd12}
\end{equation}
Here, as in the previous subsections, $\Omega = \omega_0(1 + \rho)$.
We keep  our usual scaling laws
for the small detuning $\rho = \epsilon^2R$
and damping coefficient
$\gamma = \epsilon^2\Gamma$,
but  choose a fractional order of smallness for the driving amplitude: 
$h = \epsilon^{3/2}H$.

  \begin{figure}[t]
  \center
\includegraphics{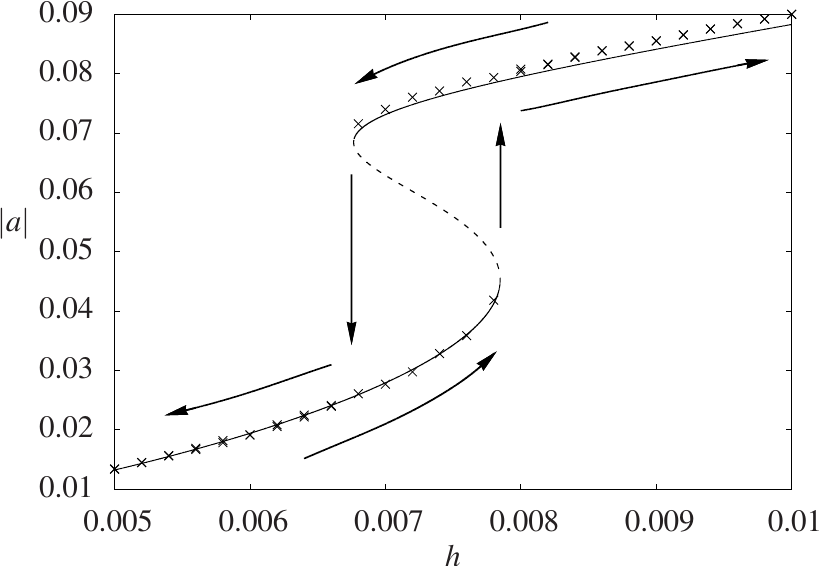}
\caption{\label{fig3} The hysteresis loop of the kink driven by an external subharmonic force.
The continuous and 
dashed lines depict the stable and unstable fixed points of the 
dynamical  system (\ref{amp3}).
Crosses result from the  numerical simulations of 
the  partial differential equation   (\ref{dd12}), with the same $\gamma$ and $\rho$.
(In this plot, $\gamma = 10^{-3}$ and $\rho = -2 \times 10^{-3}$.)
The driving strength $h$
is increased  in small  increments 
and then turned  down.
The dependence of $|a|$ on $h$ is similar to the one in Fig \ref{fig1}; yet the same driving strength sustains a much smaller wobbling amplitude
than in Fig \ref{fig1}. }
\end{figure}

One amplitude equation is standard, ${\dot v} = -2\gamma v + O(|a|^5)$.
Once the velocity has become of order $|a|^3$,  it drops out of the second equation which then acquires the form \cite{OB}
\begin{equation}
{\dot a}  +  i   \omega_0   \left(  \rho +\frac{\lambda}{2} h^2 \right)
a - \frac{i}{2}         \omega_0 \zeta              |a|^2a
= -\gamma a  +
 i \frac{60 }{169}\pi  \omega_0 h^2.
 \label{amp3}
\end{equation}
Note that the term proportional to $\lambda$ is of higher order  than all other terms in this equation. 
We had to include this forth-order correction to improve the agreement between the fixed points of the dynamical system (\ref{amp3}) and
results of simulation of the full partial differential equation (\ref{dd12}). The fact that a higher-order term makes a significant 
contribution is due to its large coefficient: $\lambda=  -7.47 -  1.68 \, i$. 

The equation (\ref{amp3}) coincides with the
equation (\ref{amp2}) that we considered earlier, with $\rho$ replaced with $\rho^\prime= \rho+\lambda h^2/2$,
 $h$  with $ h^\prime= \frac{480}{169} h^2$, and  $a^\prime= -a$.
 Hence the dynamics of the $1:2$ externally driven wobbllng kink reproduce  those of the kink driven by the $1:1$ parametric force.

As in that earlier system, 
the absence or presence of hysteresis in 
the dynamics
depends on whether $\rho^\prime$ is above or below the critical value
$\rho_0=-1.14 \gamma$. If the difference 
$\rho^\prime-\rho_0$ is positive, all trajectories are attracted to a single fixed point.
 If, on the other hand, 
$\rho^\prime-\rho_0<0$, we
have two stable fixed points
 for each $h^\prime$
in a finite interval $(h_+, h_-)$, where $h_+$ and $h_-$ are expressible through $\rho^\prime$ and $\gamma$.
This bistability  leads to the hysteretic  phenomena similar to the one depicted in Fig \ref{fig1}. 
Outside the interval $(h_+, h_-)$, all trajectories flow to a single fixed point. 
See Fig \ref{fig3}.
  
The energy transfer mechanism
associated with the $1:2$ external pumping  deserves to be commented.
It was suggested \cite{QuinteroDirect}
that  the mechanism consists in  the coupling of the wobbling to the translation   mode.
Our asymptotic expansion furnishes \cite{OB} a different explanation though.

Expanding $\phi$ in powers of $\epsilon^{1/2}$,
\[
\phi = \phi_0  + \epsilon \phi_{1} + 
\epsilon^{3/2} \phi_{3/2} + \epsilon^2 \phi_2 + \epsilon^{5/2} \phi_{5/2} + \ldots,
\]
we observe that 
 the external driver excites an 
even-parity standing wave  $\phi_{3/2}$ at its frequency $\Omega/2$.
 The standing wave 
undergoes  frequency doubling and parity 
transmutation via the term $\epsilon^3 \phi_0 \phi_{3/2}^2$ in the expansion of $\phi^3$. 
It is this  latter term that serves as an external driver for 
 the wobbling mode.  It has the resonant frequency $\Omega$ and its parity coincides with the parity of the mode.

 This energy-pumping mechanism is
a two-stage process and the resulting effective driving strength
is proportional to $h^2$ rather than $h$. 
As a result, the external subharmonic driving sustains a much smaller wobbling amplitude than the 
harmonic parametric forcing. 
This conclusion is obvious if one compares the vertical scales in Fig \ref{fig3} and Fig \ref{fig1}.

\subsection{External harmonic  driving}

Driving the kink by an external        periodic         force             with the wobbling frequency leads to the most interesting phenomenology. In this case 
the $\phi^4$ equation has the form \cite{QuinteroDirect}
\begin{equation}
\frac{1}{2}\phi_{tt}-\frac{1}{2}\phi_{xx}  - \phi +
\phi^3 = - \gamma \phi_t +
h\cos(\Omega t),
\label{dd11}
\end{equation}
where $\Omega = \omega_0(1 + \rho)$.  As in the previous three cases, we let $\rho = \epsilon^2R$
and  $\gamma = \epsilon^2\Gamma$. 
Choosing a linear scaling for the driving amplitude, 
$h = \epsilon H$, and assuming that the velocity $v=\epsilon^2 V$ 
(rather than $\epsilon V$ as before), 
we obtain the following system of two amplitude equations \cite{OB}:
\begin{equation}
\label{amp5}
{\dot a} +    i\omega_0 \left(  \rho +\frac{\nu}{2} h^2  \right)  a-  \frac{i}{2}  \omega_0 \zeta |a|^2a
= - \gamma a  
+ \frac{3\pi}{4} vh 
- \frac{i}{2} \omega_0 \mu h^2 a^* + O(|a|^5),  
\end{equation}
\begin{equation}
\label{amp6}
{\dot v} = -2\gamma v +   \frac32 h \left[  \frac{\pi}{2}  \omega_0  \left(\omega_0 \rho + i \gamma \right)
- \eta  |a|^2 
- \chi h^2
  \right]a +c.c. 
+O(|a|^5). 
\end{equation}
Here $\nu$, $\mu$, $\eta$, and $\chi$ are numerical coefficients:
\begin{eqnarray*}
\nu= 4.16 -  0.33 \, i, \quad \mu =1.02 +  0.16 \, i,
\\
\eta = -2.00 -  0.38 \, i, \quad
 \chi = -12.21 -  0.57 \, i. 
\end{eqnarray*}

\begin{figure}[t]
\center
\includegraphics{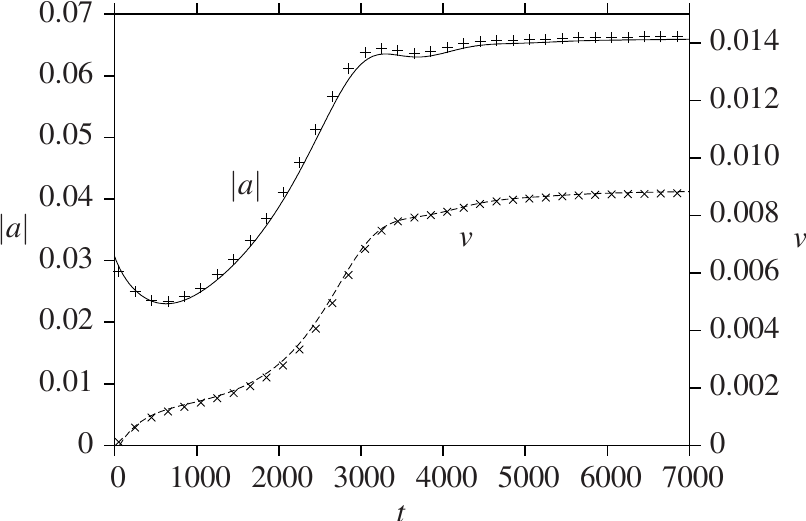}
\caption{\label{fig4} 
The wobbling kink  set in motion by
the harmonic external force. 
The curves were obtained by solving the 
amplitude equations (\ref{amp5})-(\ref{amp6}).
The crosses represent  numerical simulations of the partial differential
equation (\ref{dd11}) with the corresponding initial conditions.
In this plot,  $h = 0.012$, $\gamma = 
0.001$ and $\rho = 0$.   
}
\end{figure}

According to the equation (\ref{amp6}), the velocity does not drop out of the system; instead, it plays an important role in the dynamics. 
Unlike all types of driving that we considered so far, 
the harmonic external driving can sustain the translational motion
of the kink. Fig \ref{fig4} shows an example of
the kink accelerated by the 1:1 direct driving force which
simultaneously excites the wobbling \cite{OB}.

\begin{figure}[t]
\includegraphics[width=0.49\linewidth]{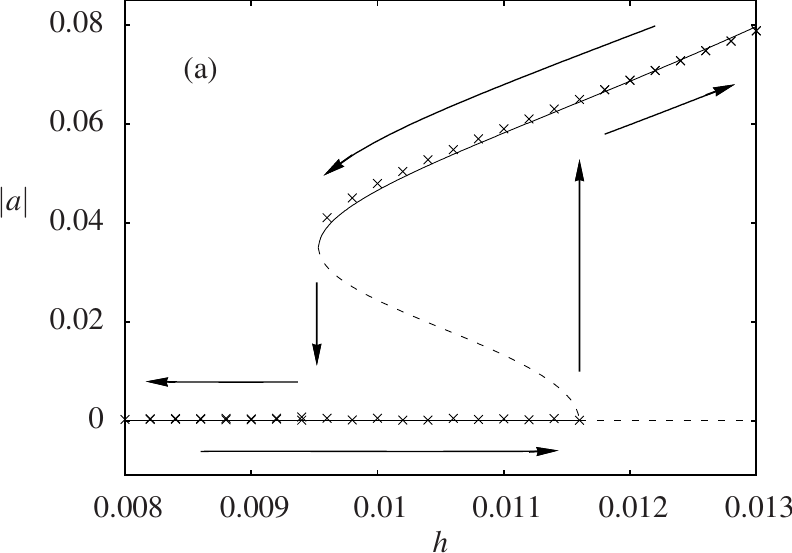}
\hspace*{3mm}
 \includegraphics[width=0.49\linewidth]{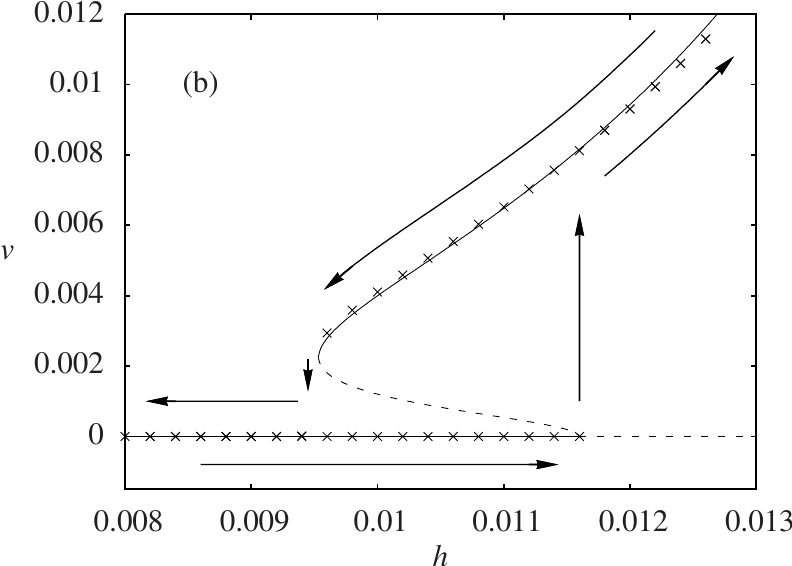}
\caption{\label{fig5}
The hysteresis loop  of the wobbling kink driven by an external harmonic force.
The continuous and dashed curves delineate the 
stable and unstable fixed points of the amplitude equations
(\ref{amp5})-(\ref{amp6}),
with $\gamma =   10^{-3}$ and $\rho=-  10^{-4}$.
The crosses
indicate results of the  numerical simulations of the  
partial differential equation (\ref{dd11})  with the same $\gamma$ and $\rho$. 
The driving amplitude  is increased from 
$h=8 \times 10^{-3}$ to $13 \times 10^{-3}$ in
small steps and then turned back to $8 \times 10^{-3}$.}
\end{figure}

When $h$ is small, the fixed point at $a=v=0$ is the only attractor in the system.
As $h$ is increased while keeping the parameters $\gamma$ and $\rho$ constant, a pair of  fixed points $(a,v)$ and $(-a,-v)$, with nonzero $a$ and $v$, 
bifurcates from the trivial fixed point 
(see Fig.\ref{fig5}).

 As the driving strength approaches some critical
 value $h_{\rm c}$,  the 
$|a|$- and $v$-component of these points grow to infinity. 
No stable fixed
points exist in the parameter region beyond $h_{\rm c}$.
In that region, all trajectories escape to infinity:
$|a(t)| \to \infty, v(t) \to \infty$ as $t \to \infty$.
The critical value $h_{\rm c}$ can be determined assuming that the growth is self-similar,
that is, that $v$ grows as a power of $|a|$.  This assumption leads to a simple exponential asymptote
$e^{\frak r t}$  for $|a(t)|$, with the growth rate \cite{OB}
\[
{\frak r}= -\frac23 \gamma - \frac{3 \pi i }{4 \omega_0} \,
\frac{\eta \zeta^*-\zeta \eta^*}{|\zeta|^2}
 h^2.
\]
Setting $\frak r=0$ gives
\begin{equation}
h_{\rm c}=0.65 \gamma^{1/2}.
\label{threshold}
\end{equation}

In the vicinity of  the critical value (\ref{threshold})
our smallness assumptions about $|a|$ and $v$ are no longer met
and the amplitude system (\ref{amp5})-(\ref{amp6}) is no longer valid.
The simulations of the full partial differential equation (\ref{dd11}) with
 $h$ just below and just above $h_{\rm c}$,
 reveal that the kink does settle to the wobbling with finite amplitude here.
 This is accompanied by its motion with constant velocity.
 However, the observed value of the wobbling amplitude is $O(\gamma^{1/3})$ 
 rather than $O(\gamma^{1/2})$ as we assumed in the derivation of  (\ref{amp5})-(\ref{amp6}),
 and the measured value of the kink velocity is $O(\gamma^{1/3})$ rather than $O(\gamma)$.
 This change of scaling accounts for the breakdown of our asymptotic expansion in the vicinity of $h_{\rm c}$.

Since the spatially uniform (even) driving force has an opposite parity to that of the (odd) wobbling mode, 
the energy transfer mechanism has to be indirect here. In fact there are two mechanisms at work,
both involving a standing wave with the frequency $\Omega$
excited by the external driver. In the first mechanism, the square of 
this wave, its 
zeroth and second harmonics as well as the radiation excited by this wave, couple, parametrically, to the wobbling mode.
The second mechanism exploits the fact that 
when the kink moves relative to the 
standing wave, the  wobbling mode acquires an
even-parity component. It is this part of the mode that 
 couples --- directly --- to the standing wave.

\section{Concluding remarks}
\label{summary}

The first objective of this article was to review 
  the multiscale singular perturbation
expansion for the wobbling kink of the  $\phi^4$ model. The advantage of this approach over 
the  Lindstedt method, is that it
takes into account the existence of a hierarchy of  space and time scales in
the system.  In particular, 
  the multiscale expansion provides a consistent treatment  of the
long-range radiation from the oscillating kink. 

The central outcome of the asymptotic analysis is the 
equation  (\ref{mainampeqfree})
  for the amplitude of the wobbling mode.    The nonlinear frequency shift and   the lifetime  of the wobbling mode are straightforward from 
this   amplitude equation.
We have identified the main factor  ensuring its longevity.
This is a small amplitude of radiation due to
the significant difference between the wavelength of the radiation
and the characteristic width of the wobbling kink.

The second part of this brief review (section \ref{drive})
concerned ways of sustaining the wobbling of the kink indefinitely.
We discussed four resonant frequency regimes, namely the $1:1$ and $2:1$ parametric driving, 
and $1:2$ and $1:1$ external forcing.

It is instructive to compare the amplitude of the stationary wobbling mode
sustained by these 
four types of resonance. 
For the given driving strength $h$,  
the  harmonic  ($1:1$)  parametric driver ensures the strongest possible response.
In this
case the amplitude of the stationary kink oscillations 
is of the order  of $h^{1/3}$. The ``standard"  parametric resonance, where the 
driving frequency is chosen to be near twice the natural wobbling frequency of the kink,  is 
second strongest.  In this case 
 the wobbling amplitude is of the order of $h^{1/2}$.
 The direct driving at half the wobbling frequency produces oscillations with the amplitude
$a \sim h^{2/3}$. Finally, the harmonic   ($1:1$) direct resonance is the weakest  of the four responses considered.
In this case, the amplitude of the forced oscillations of the kink is
$a \sim h$. 

Another resonance characteristic  worth comparing, is 
the width of the ``Arnold tongue" --- the domain on the  $(\Omega, h)$ parameter  plane where the driver  sustains stable stationary wobbling of the kink. 
The $1:1$  parametric driver produces  the widest tongue; in this
case the resonant region is bounded by the curve $h \sim \rho^{3/2}$. 
The ``standard" ($2:1$) parametric resonance is the second widest, bounded by 
the curve $h \sim \rho$. The Arnold tongue for the external driving force with the frequency $\Omega/2$,
 is bounded by
$h \sim \rho^{3/4}$.
 Finally,  the harmonic  direct resonance is the narrowest of the four, with the boundary curve
$h \sim \rho^{1/2}$.

Our comparison would be incomplete without noting that 
neither 
the harmonic parametric driver nor
 the  $\Omega/2$ direct driving force  have to overcome any thresholds
in order to sustain stationary wobbling 
with a nonzero amplitude.
 In contrast,
 the subharmonic ($2:1$) parametric and harmonic ($1:1$) direct resonances occur
only if the driving amplitude exceeds a certain threshold value.

\begin{figure}
\center
 \includegraphics[width=0.7\linewidth]{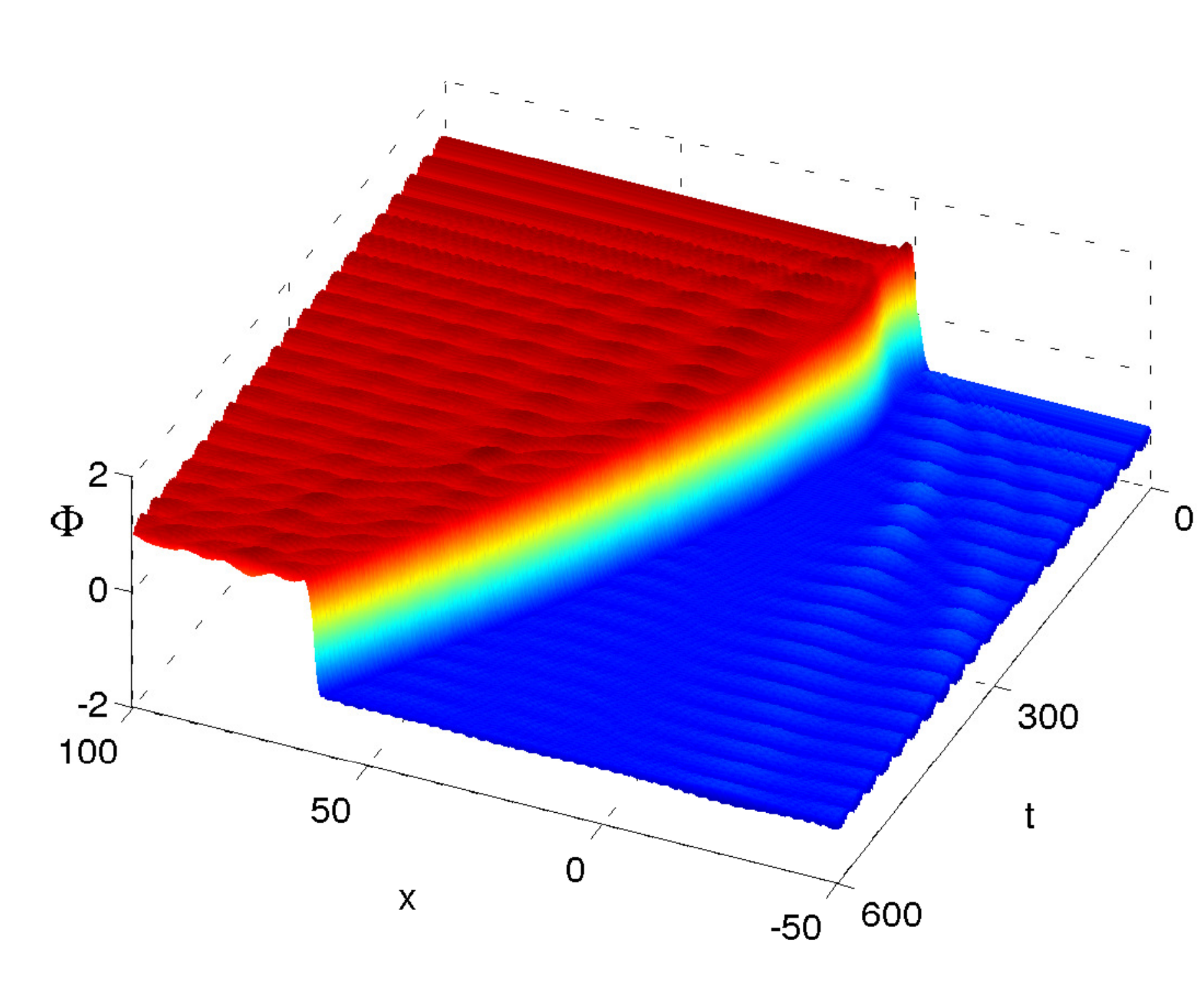}
\caption{A  kink accelerates from rest as  its wobbling mode couples to the standing wave excited by the 
resonant harmonic force. 
The figure is produced by numerical simulation of equation \eqref{dd11} with $\gamma=10^{-3}$, $\rho=10^{-2}$, and $h= 4 \times 10^{-2}$. 
The initial conditions are $\phi_t=0$ and $\phi= \tanh x+ 2 a  \tanh x  \, \mathrm{sech} \, x$  with $a=0.3$.
}
\label{translation}       
\end{figure}

Thus, the harmonic external driving  emerges as the least efficient way of sustaining the steady wobbling of the kink.
Out of the four driving techniques considered in section \ref{drive}, 
it produces the weakest response, requires the finest tuning of the driving frequency while the corresponding driving strength has to overcome 
a  threshold set by the dissipation.
Although these factors are indeed disadvantageous, the harmonic external driving gives rise to an interesting ``rack and pinion" mechanism
that converts the energy of external oscillation to the translational motion of the kink (Fig \ref{translation}). 
This mechanism may prove  useful for the control of solitons with internal modes in other systems.

  \vspace*{-5mm}

\section{Epilogue}
This brief review is a tribute to Boris Getmanov, a musician and nonlinear scientist.
A whole generation of former  school kids still recalls gyrating to his band's boogie grooves
while those with a taste for integrable systems remember Getmanov's discovery of the  complex sine-Gordon  \cite{BG1}.
(Incidentally, that discovery  was only possible due to his numerical experimentation with the $\phi^4$ model.)
While Bob's piano is no longer heard at Dubna  parties, the $\phi^4$ and sine-Gordon equations are still around, 
alluring their new insomniacs and artists.

\section*{Acknowledgments}
Most of the  results reviewed in this chapter were obtained jointly with 
 Oliver Oxtoby \cite{OB,BO}.
 I am grateful to Oliver
 for his collaboration on the wobbling kink project.
 Special thanks go to Nora Alexeeva  for generating  Figs \ref{wobble}, \ref{breakup} and \ref{translation}  for this piece.


\printindex
\end{document}